# Reliability Analysis of a 1-out-of-*n* Cold Standby Redundant System under the Generalized Lindley Distribution

Afshin Yaghoubi, Ph.D. candidate

Faculty of Mathematics and Computer Science, Amirkabir University of Technology, Tehran, Iran;

Email: afshin.y@aut.ac.ir

Esmaile Khorram, Ph.D.

Faculty of Mathematics and Computer Science, Amirkabir University of Technology, Tehran, Iran;

Email: eskhor@aut.ac.ir

Omid Naghshineh Arjmand[1], Ph.D.

Faculty of Mathematics and Computer Science, Amirkabir University of Technology, Tehran, Iran;

Email: naghshineh@aut.ac.ir

**Abstract**

Cold standby 1-out-of-*n* redundant systems are well-established models in system reliability engineering. To date, reliability analyses of such systems have predominantly assumed exponential, Erlang, or Weibull failure distributions for their components. The Lindley distribution and its generalizations represent a significant class of statistical distributions in reliability engineering. Certain generalized Lindley distributions, due to the appealing characteristics of their hazard functions, can serve as suitable alternatives to other well-known lifetime distributions like the Weibull. This study investigates the reliability of a 1-out-of-*n* cold standby redundant system with perfect and imperfect switching, assuming that the active component failure times follow the Generalized Lindley distribution. We derive a closed-form expression for the system reliability. To achieve this, the distribution of the sum of *n* independent and identically distributed random variables following the Generalized Lindley distribution is first determined using the moment-generating function approach.

---

[1] Corresponding Author

1. **Introduction**

Technological advancements across all fields have compelled engineers and system designers to rigorously evaluate the concept of reliability, as it is a vital tool for enhancing productivity, efficiency, and product quality. With the rapid growth of technology in various domains, the complexity and automation of systems are increasing exponentially. Standby redundant systems are one such design proposed by engineers. From an energy perspective, the most ideal system for minimizing costs is the 1-out-of-*n* cold standby system, as only one component is operational at any given time during system function [1]. The design of such systems requires a significant budget, which is naturally justified by the critical missions they are entrusted with.

The cold standby strategy is a widely used redundancy technique that has attracted considerable research attention. In this mode, standby components are completely shut down and unloaded, resulting in a failure rate of zero. The primary application of cold standby is typically found in areas with limited energy or electrical resources [1]. Examples include digital and electronic systems, textile production lines, exploration and detonation systems, aerospace-related satellite systems, n-modular redundant systems, and power systems such as power supplies, among others [2, 3].

The operational mechanism of a 1-out-of-*n* cold standby redundant system is interpreted as follows: Let $T_i$ denote the time to failure of component $A_i$ in its active state. If $A_1$ fails at time $T_1$, $A_2$ is immediately activated via a switch to replace $A_1$ and maintain system operation. If $A_2$ fails at time $T_1 + T_2$, $A_3$ is activated and switched in to take over system functionality. This process continues until the last redundant component is brought online [3].

The switch plays a crucial role in such systems. Its function is to continuously monitor the health of the active component and, upon its failure, immediately detect the fault and activate the standby component to restore system operation. Researchers often model the switch as either perfect or imperfect. A perfect switch is completely reliable throughout the mission, whereas an imperfect switch has a non-zero failure rate; it may fail during its operation. Two common scenarios are typically defined for an imperfect switch: 1) switch failure as a continuous function of time, and 2) switch failure only during the switching instant [4].

Common assumptions made for mathematical simplicity in the reliability analysis of these systems typically include [1-3]:

- Standby components in the cold state are perfectly healthy.
- The switching time is negligible.
- The standby components are activated in a predetermined sequence.

The reliability analysis of 1-out-of-*n* cold standby systems (CSS) was initially explored for perfect switches and exponential component failure distributions in foundational reliability textbooks such as those by [5-7]. This system analyzed with an imperfect switch, assuming component failure times followed an Erlang distribution [4]. The [8] computed the reliability of the aforementioned system of [4] using a Markov approach. The [9] reported the 1-out-of-$n$ CSS with an imperfect switch, where active component failures followed an exponential distribution, and the switch failure rate, also exponential, increased with each switching operation. The [2] approximated the reliability of CSS with a perfect switch for arbitrary component lifetime distributions using a method based on the Central Limit Theorem. Subsequently, the [3] investigated the same system with non-identical components. The [10] evaluated the reliability of 1-out-of-$n$ CSS with a perfect switch, where the active component failure times followed a Weibull distribution. They derived a system reliability equation using polynomial and Maclaurin series expansions.

The Lindley distribution is a significant statistical distribution in reliability and lifetime data modeling, introduced by [11] in 1958. It was extensively examined by [12], who demonstrated its greater flexibility compared to the exponential distribution. Subsequently, various generalizations of the Lindley distribution have been proposed. Notable examples include the generalized Lindley distribution (GLD) by [13], the GLD by [14], the GLD by [15], and the GLD by [16]. Beyond lifetime data modeling, GLD's have also found applications in energy studies. For instance, while distributions such as Weibull, Rayleigh, and log-normal are commonly used to estimate wind speed parameters, with the Weibull often providing a good fit, but [17, 18] employed a new GLD and demonstrated its superior fit compared to the Weibull, Rayleigh, and log-normal distributions. However, the application of GLD's distribution in system reliability remains largely unexplored. This paper aims to address this gap.

The authors in [15] claim that their proposed distribution can serve as an alternative to the Weibull, Gamma, and log-normal distributions. Its probability density function (PDF) possesses attractive properties, enabling it to describe strictly decreasing, strictly increasing, and bathtub-shaped hazard rates. Furthermore, since the distribution proposed in [16] encompasses other aforementioned generalizations, it is selected for this study. Therefore, this paper provides a detailed report on the reliability of a 1-out-of-$n$ cold standby system where the active component failure times follow GLD of the [16].

The organization of the paper is as follows: Section 2 introduces the generalized Lindley distribution. Section 3 presents the reliability analysis of a 1-out-of-$n$ cold standby system under perfect and imperfect switching. Section 4 provides the sensitivity analysis. Numerical results are discussed in Section 5, and conclusions along with future recommendations are given in Section 6.

2. **The Generalized Lindley Distribution**

Let $T_i \sim GLD(\alpha, \beta, \theta, \gamma, \eta)$, the PDF of it with parameters $\alpha, \beta, \theta > 0, \gamma, \eta \geq 1$ is given by:

$$f_{GLD}(t) = \frac{\gamma \theta^\alpha t^{\alpha-1}}{(\gamma + \theta^\eta)\Gamma(\alpha)} e^{-\theta t} + \frac{\theta^{\beta+\eta} t^{\beta-1}}{(\gamma + \theta^\eta)\Gamma(\beta)} e^{-\theta t}, t \geq 0 \qquad (1)$$

The Eq. (1) was introduced by [16] and is constructed from the convex combination of the $G(\alpha, \theta)$ distribution with $p_1 = \frac{\gamma}{\gamma+\theta^\eta}$ and the $G(\beta, \theta)$ distribution with $p_2 = \frac{\theta^\eta}{\gamma+\theta^\eta}$.

By selecting appropriate parameters, the generalized distributions proposed in [13-15] can be obtained. As well, in Equation (1), when $\alpha, \beta = 1$, the Exponential distribution is derived, and when $(\alpha, \beta, \theta, \gamma, \eta) = (2,1, \theta, 1,1)$, it yields the one-parameter Lindley distribution.

The reliability function of Eq. (1) is given by:

$$R_{GLD}(t) = P(X > t) = \frac{\gamma \theta^\alpha}{(\gamma + \theta^\eta)\Gamma(\alpha)} \int_t^\infty x^{\alpha-1} e^{-\theta x} dx + \frac{\theta^{\beta+\eta}}{(\gamma + \theta^\eta)\Gamma(\beta)} \int_t^\infty x^{\beta-1} e^{-\theta x} dx$$

$$= \frac{\gamma}{(\gamma + \theta^\eta)\Gamma(\alpha)} \int_{\theta t}^\infty u^{\alpha-1} e^{-u} du + \frac{\theta^\eta}{(\gamma + \theta^\eta)\Gamma(\beta)} \int_{\theta t}^\infty u^{\beta-1} e^{-u} du$$

Each of the above integrals represents an upper incomplete Gamma function, in other words, $\int_{\theta t}^\infty u^{\alpha-1} e^{-u} du = \Gamma(\alpha, \theta t)$, and $\int_{\theta t}^\infty u^{\beta-1} e^{-u} du = \Gamma(\beta, \theta t)$. Therefore, $R_{GLD}(t)$ is equals:

$$R_{GLD}(t) = \frac{\gamma}{\gamma + \theta^\eta} \left(\frac{\Gamma(\alpha, \theta t)}{\Gamma(\alpha)}\right) + \frac{\theta^\eta}{\gamma + \theta^\eta} \left(\frac{\Gamma(\beta, \theta t)}{\Gamma(\beta)}\right), t \geq 0. \qquad (2)$$

In Eq. (2), when $\alpha$ and $\beta$ are positive integer, then $\Gamma(\alpha, \theta t)$ and $\Gamma(\beta, \theta t)$ are given by

$$\Gamma(\alpha, \theta t) = (\alpha - 1)! \, e^{-\theta t} \sum_{k=0}^{\alpha-1} \frac{(\theta t)^k}{k!}, \Gamma(\beta, \theta t) = (\beta - 1)! \, e^{-\theta t} \sum_{k=0}^{\beta-1} \frac{(\theta t)^k}{k!}.$$

Nevertheless, for $\theta > 0, \gamma, \eta \geq 1$, $R_{GLD}(t)$ reduces to:

$$R_{GLD}(t) = \left(\frac{\gamma}{\gamma + \theta^\eta} \sum_{k=0}^{\alpha-1} \frac{(\theta t)^k}{k!} + \frac{\theta^\eta}{\gamma + \theta^\eta} \sum_{k=0}^{\beta-1} \frac{(\theta t)^k}{k!}\right) e^{-\theta t}, t \geq 0.$$

The hazard rate function of Eq. (1), for $\alpha, \beta, \theta > 0, \gamma, \eta \geq 1$, is obtained as follows:

$$h_{GLD}(t) = \frac{f_{GLD}(t)}{R_{GLD}(t)} = \frac{\gamma \frac{\theta^{\alpha}}{\Gamma(\alpha)} t^{\alpha-1} + \theta^{\eta} \frac{\theta^{\beta}}{\Gamma(\beta)} t^{\beta-1}}{\gamma \frac{\Gamma(\alpha,\theta t)}{\Gamma(\alpha)} + \theta^{\eta} \frac{\Gamma(\beta,\theta t)}{\Gamma(\beta)}} e^{-\theta t}, t \geq 0 \qquad (3)$$

The hazard function, having five parameters, can be show various behaviors. For example, when $\alpha, \beta = 1$, then $\Gamma(1, \theta t) = e^{-\theta t}$, so $h_{GLD}(t) = \theta$. when $\alpha, \beta = 2$, then $\Gamma(2, \theta t) = (1 + \theta t)e^{-\theta t}$, in this case, $h_{GLD}(t) = \frac{\theta^2 t}{1+\theta t}$, thus can be simply proved that $h_{GLD}(t)$ is increasing.

We know that the gamma density function $G(k, \theta)$ for $k = \alpha, \beta$, possess the log-concavity property for $k > 1$, because $\left(log\left(\frac{\theta^k t^{k-1} e^{-\theta t}}{\Gamma(k)}\right)\right)'' = \frac{1-k}{t^2} < 0$. Since $f_{GLD}(t)$ is a convex combination of these two of the $G(\alpha, \theta)$ and the $G(\beta, \theta)$ distribution and satisfy $\alpha, \beta > 1$, the resulting mixture remains log-concave [19, 20]. According to fundamental reliability theory, log-concavity of the density implies a strictly increasing hazard rate [20]. Consequently, for $\alpha, \beta > 1$, the $h_{GLD}(t)$ is strictly increasing.

Similarly, the log-convexity property for $k < 1$, because $\left(log\left(\frac{\theta^k t^{k-1} e^{-\theta t}}{\Gamma(k)}\right)\right)'' = \frac{1-k}{t^2} > 0$, Hence, for $\alpha, \beta < 1$, the $h_{GLD}(t)$ is strictly decreasing.

When $\alpha < 1, \beta > 1$ or $\alpha > 1, \beta < 1$, the $h_{GLD}(t)$ exhibits a U-shaped (bathtub) behavior. This occurs because:

- As $t \to 0$, $h_{GLD}(t) \to \infty$ due to the $t^{\alpha-1}$ term with $\alpha < 1$ dominating.
- As $t \to \infty$, $h_{GLD}(t) \to 0$ because of the exponential decay $e^{-\theta t}$.
- The derivative $h_{GLD}(t)$ changes sign exactly once, creating a unique minimum point, which follows from the monotonicity properties of the two gamma components.

Hence, $h_{GLD}(t)$ decreases from infinity to a minimum, then increases again, forming the characteristic U-shape.

The graph of the hazard function in Eq. (3), for some different parameters, is shown in Fig. 1.

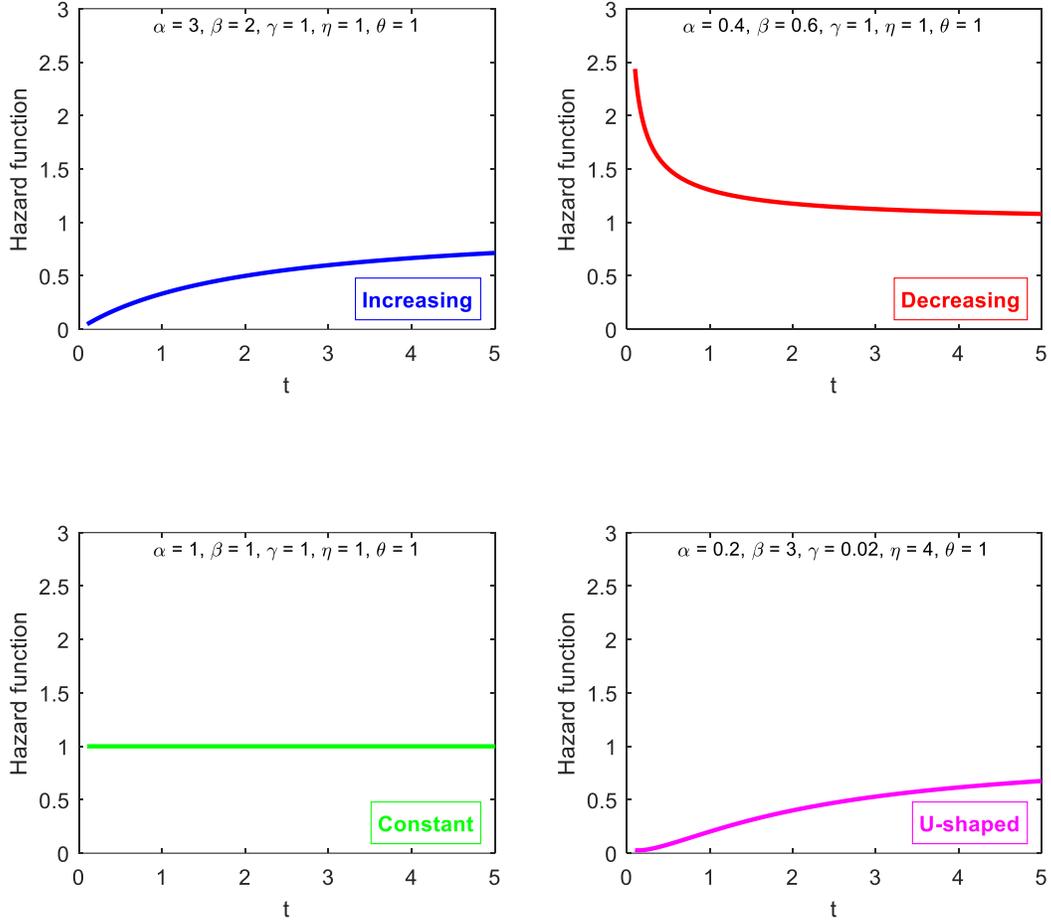

**Fig. 1. Hazard function plot for different parameters.**

In the above figures, it is observed that Eq. (3) models strictly decreasing, strictly increasing, constant and U-shaped hazard functions.

*2.1. Distribution of the Sum of n iid Generalized Lindley Variables*

**Theorem 1.** Let $X_1, X_2, \ldots, X_n$ be a random sample from Eq. (1). The PDF of the sum of *n iid* random variables from Eq. (1), for $\alpha, \beta, \theta > 0, \gamma, \eta \geq 1$, using the moment generating function (MGF), is equal to:

$$f_{S_n}(x) = \left(\frac{\theta^\eta}{\gamma + \theta^\eta}\right)^n \sum_{i=0}^{n} \binom{n}{i} (\gamma\theta^{-\eta})^i \left(\frac{\theta^{i\alpha+(n-i)\beta} x^{i\alpha+(n-i)\beta-1}}{\Gamma(i\alpha + (n-i)\beta)} e^{-\theta x}\right), x \geq 0 \qquad (4)$$

**Proof.** The MGF of Eq. (4) is determined as follows:

$$M_X(t) = E(e^{tX}) = \frac{\gamma}{\gamma + \theta^\eta} \int_0^\infty e^{tx} f_G(x; \alpha, \theta) dx + \frac{\theta^\eta}{\gamma + \theta^\eta} \int_0^\infty e^{tx} f_G(x; \beta, \theta) dx$$

Each of the integrals above is the moment generating function of the Gamma distribution. It is well known that the moment generating function of a Gamma distribution with shape parameter $\alpha$ and scale parameter $\theta$ is $\left(1 - \frac{t}{\theta}\right)^{-\alpha}$. Therefore, the moment generating function of the above equation is:

$$M_X(t) = \left(\frac{\gamma}{\gamma + \theta^\eta}\right)\left(1 - \frac{t}{\theta}\right)^{-\alpha} + \left(\frac{\theta^\eta}{\gamma + \theta^\eta}\right)\left(1 - \frac{t}{\theta}\right)^{-\beta}$$

Since the $X_i$ are independent and identically distributed, the moment generating function of their sum, $S_n = X_1 + \cdots + X_n$, is:

$$M_{S_n}(t) = (M_X(t))^n = \left(\left(\frac{\gamma}{\gamma + \theta^\eta}\right)\left(1 - \frac{t}{\theta}\right)^{-\alpha} + \left(\frac{\theta^\eta}{\gamma + \theta^\eta}\right)\left(1 - \frac{t}{\theta}\right)^{-\beta}\right)^n$$

Using the binomial expansion, $M_{S_n}(t)$ equals:

$$M_{S_n}(t) = \sum_{i=0}^{n} \binom{n}{i} \left(\frac{\gamma}{\gamma + \theta^\eta}\right)^i \left(\frac{\theta^\eta}{\gamma + \theta^\eta}\right)^{n-i} \left(1 - \frac{t}{\theta}\right)^{-(i\alpha+(n-i)\beta)}$$

It can be readily seen that $\left(1 - \frac{t}{\theta}\right)^{-(i\alpha+(n-i)\beta)}$ is the moment generating function of a Gamma distribution with shape parameter $i\alpha + (n-i)\beta$ and scale parameter $\theta$. Hence, the proof is complete.

3. **Reliability of a 1-out-of-$n$ Cold Standby Redundant System**

In [4], it was shown that the reliability of a 1-out-of-$n$ cold standby redundant system with a perfect switch is equal to:

$$R_{Sys.}(t)_{Perfect} = r(t) + \sum_{i=1}^{n-1} \int_0^t f_{S_i}(u) r(t-u) \, du \qquad (5)$$

where $n$ is the number of components in the system, and $r(.), f_{S_i}(.)$ are the reliability of the active component and the probability density function of the sum of $i$ components, respectively. If $f_{S_i}(.)$ is available in closed form, the task becomes straightforward. Under these conditions, through a series of direct algebraic calculations, the reliability can be computed as an explicit equation.

*3.1. System Reliability with a Perfect Switch*

According to Theorem 1, and using Eqs. (2) and (4), the system reliability under a perfect switch can be calculated with some algebraic computations as follows:

$$R_{Sys.}(t)_{Perfect}^{GLD}$$
$$= \frac{\gamma}{\gamma + \theta^\eta} e^{-\theta t} \sum_{i=0}^{n-1} \sum_{j=0}^{i} \sum_{k=0}^{\alpha-1} \binom{i}{j} \left(\frac{\gamma}{\theta^\eta}\right)^j \left(\frac{\theta^\eta}{\gamma + \theta^\eta}\right)^i \frac{(\theta t)^{j\alpha + (i-j)\beta + k}}{(j\alpha + (i-j)\beta + k)!}$$
$$+ \frac{\theta^\eta}{\gamma + \theta^\eta} e^{-\theta t} \sum_{i=0}^{n-1} \sum_{j=0}^{i} \sum_{k=0}^{\beta-1} \binom{i}{j} \left(\frac{\gamma}{\theta^\eta}\right)^j \left(\frac{\theta^\eta}{\gamma + \theta^\eta}\right)^i \frac{(\theta t)^{j\alpha + (i-j)\beta + k}}{(j\alpha + (i-j)\beta + k)!} \quad (6)$$

By substituting $\alpha, \beta = 1$ into Eq. (6), the system reliability under the exponential distribution is simply obtained as follows:

$$R_{Sys.}(t)_{Perfect}^{E} = e^{-\theta t} \sum_{i=0}^{n-1} \frac{(\theta t)^i}{i!} \quad (7)$$

This relationship is available in [5-7]. Furthermore, if the parameters $(\alpha, \beta, \theta, \gamma, \eta) = (2, 1, \theta, 1, 1)$ are substituted into Eq. (6), the system reliability under the Lindley distribution is obtained, after some straightforward calculations, as follows:

$$R_{Sys.}(t)_{Perfect}^{L} = e^{-\theta t} \sum_{i=0}^{n-1} \sum_{j=0}^{i} \left(\frac{\theta^2}{1+\theta}\right)^i \binom{i}{j} \frac{t^{i+j}}{(i+j)!} \left(1 + \frac{t}{i+j+1}\right) \quad (8)$$

The mean time to failure of the system (*MTTF*) under the GLD can be calculated as follows:

$$MTTF_{Perfect}^{GLD} = \int_0^\infty R_{Sys.}(t)_{Perfect}^{GLD} dt$$
$$= \frac{\gamma}{\gamma + \theta^\eta} \sum_{i=0}^{n-1} \sum_{j=0}^{i} \sum_{k=0}^{\alpha-1} \binom{i}{j} \left(\frac{\gamma}{\theta^\eta}\right)^j \left(\frac{\theta^\eta}{\gamma + \theta^\eta}\right)^i \frac{\theta^{j\alpha + (i-j)\beta + k}}{(j\alpha + (i-j)\beta + k)!} \int_0^\infty t^{j\alpha + (i-j)\beta + k} e^{-\theta t} dt$$

$$+\frac{\theta^\eta}{\gamma+\theta^\eta}\sum_{i=0}^{n-1}\sum_{j=0}^{i}\sum_{k=0}^{\beta-1}\binom{i}{j}\left(\frac{\gamma}{\theta^\eta}\right)^j\left(\frac{\theta^\eta}{\gamma+\theta^\eta}\right)^i\frac{\theta^{j\alpha+(i-j)\beta+k}}{(j\alpha+(i-j)\beta+k)!}\int_0^\infty t^{j\alpha+(i-j)\beta+k}e^{-\theta t}dt \quad (9)$$

Simplicity can be proved that

$$\int_0^\infty t^{j\alpha+(i-j)\beta+k}e^{-\theta t}dt = \frac{(j\alpha+(i-j)\beta+k)!}{\theta^{j\alpha+(i-j)\beta+k+1}}$$

So, Eq. (9) can be expressed as

$$MTTF_{Perfect}^{GLD} = \frac{\gamma\alpha}{\theta(\gamma+\theta^\eta)}\sum_{i=0}^{n-1}\left(\frac{\theta^\eta}{\gamma+\theta^\eta}\right)^i\sum_{j=0}^{i}\binom{i}{j}\left(\frac{\gamma}{\theta^\eta}\right)^j$$

$$+\frac{\beta\theta^\eta}{\theta(\gamma+\theta^\eta)}\sum_{i=0}^{n-1}\left(\frac{\theta^\eta}{\gamma+\theta^\eta}\right)^i\sum_{j=0}^{i}\binom{i}{j}\left(\frac{\gamma}{\theta^\eta}\right)^j$$

Using the binomial expansion, we know very well that $\sum_{j=0}^{i}\binom{i}{j}\left(\frac{\gamma}{\theta^\eta}\right)^j = \left(\frac{\gamma+\theta^\eta}{\theta^\eta}\right)^i$. Therefore, *MTTF* reduces to

$$MTTF_{Perfect}^{GLD} = \frac{n}{\theta}\left(\frac{\gamma\alpha+\beta\theta^\eta}{\gamma+\theta^\eta}\right) \quad (10)$$

By substituting the appropriate parameters into Eq. (10), the *MTTF* of the system under the exponential and Lindley distributions are determined from Eqs. (11) and (12), respectively:

$$MTTF_{Perfect}^{E} = \frac{n}{\theta} \quad (11)$$

$$MTTF_{Perfect}^{L} = \frac{n}{\theta}\left(\frac{2+\theta}{1+\theta}\right) \quad (12)$$

*3.2. Reliability of a 1-out-of-n Cold Standby System with an Imperfect Switch*

The reliability of a 1-out-of-*n* cold standby redundant system with an imperfect switch is defined as [4]:

$$R_{Sys.}(t)_{Imperfect} = r(t) + \sum_{i=1}^{n-1}\int_0^t r_s(u)f_{S_i}(u)r(t-u)\,du \quad (13)$$

where $r_s(.)$ represents the switch reliability.

For the case where the active component failures follow an Erlang distribution and the switch failure time follows an exponential distribution, a lower bound for the system reliability was proposed by [4]. Here, it is assumed that failures of the active components and the switch follow

Eq. (1). Since exact calculations become very difficult, we also seek a lower bound for the system reliability. Given that

$$r_s(u) = \left( \frac{\gamma_s}{\gamma_s + \theta_s^{\eta_s}} \sum_{k=0}^{\alpha_s-1} \frac{(\theta_s u)^k}{k!} + \frac{\theta_s^{\eta_s}}{\gamma_s + \theta_s^{\eta_s}} \sum_{k=0}^{\beta_s-1} \frac{(\theta_s u)^k}{k!} \right) e^{-\theta_s u},$$

is a decreasing function of $u$, the inequality $r_s(t) \leq r_s(u)$ always holds for $u \leq t$. Accordingly, a lower bound for the system reliability under an imperfect switch is defined as $\hat{R}_{Sys.}(t)_{Imperfect}$ and, according to Eq. (13), is calculated as:

$$\hat{R}_{Sys.}(t)_{Imperfect} = r(t) + r_s(t) \sum_{i=1}^{n-1} \int_0^t f_{S_i}(u) r(t-u)\, du.$$

According to Eq. (5), $\sum_{i=1}^{n-1} \int_0^t f_{S_i}(u) r(t-u)\, du$ is equal to:

$$\sum_{i=1}^{n-1} \int_0^t f_{S_i}(u) r(t-u)\, du = R_{Sys.}(t)_{Perfect} - r(t).$$

Therefore, the reliability of the mentioned system under an imperfect switch is equal to:

$$\hat{R}_{Sys.}(t)_{Imperfect} = r(t) + r_s(t) \left( R_{Sys.}(t)_{Perfect} - r(t) \right). \tag{14}$$

It is evident that in Eq. (14), when $r_s(t) = 1$, then $\hat{R}_{Sys.}(t)_{Imperfect} = R_{Sys.}(t)_{Perfect}$.

4. **Sensitivity Analysis**

Sensitivity analysis is a highly useful tool for decision-makers and managers. Recently, some researchers have also applied sensitivity analysis in the field of system reliability. Sensitivity analysis is a method for systematically varying the inputs of a statistical model to predict the effects of these changes on the model's output. In other words, it examines how changes in the input parameters of the model will affect the output parameter. The computational approach to sensitivity analysis is based on partial derivatives. Various works have been conducted in the field of system reliability; for example, references [21-23] can be mentioned.

Here, sensitivity analysis is performed on the system reliability and *MTTF* (i.e., Eqs. (6) and (10)). The sensitivity analysis of the system reliability function and the *MTTF* index are determined by the relations $\frac{\partial R_{Sys.}(t)_{Perfect}^{GLD}}{\partial \Theta}$ and $\frac{\partial MTTF_{Perfect}^{GLD}}{\partial \Theta}$, respectively, where $\Theta = \{\alpha, \beta, \theta, \gamma, \eta\}$.

5. **Numerical Example**

To bridge the gap between theoretical formulation and practical implementation, this section employs a canonical industrial case: the cold standby 1-out-of-*n* architecture found in aircraft emergency braking systems [24, 25]. A numerical analysis is conducted by varying the system's redundancy level (*n* = 2, 5, 10, 20) to assess the scalability and effectiveness of the proposed model under different design configurations. A mission time of 100 hours is assumed for the system. The failure time of the system components in the active state follows the GLD with ($\alpha = 2, \beta = 3, \theta = 0.5, \gamma = 1.5, \eta = 2.2$). The switch failure time is assumed to follow the same distribution with ($\alpha_s = \beta_s = 4, \theta_s = 0.005, \gamma_s = \eta_s = 1$). According to Equation 2, the closer $\theta_s$ is to zero, the higher the reliability. Therefore, the parameters of the switch failure time are set such that the switch is more reliable than the system components. Fig. 2 shows the system reliability under perfect and imperfect switches for different numbers of components.

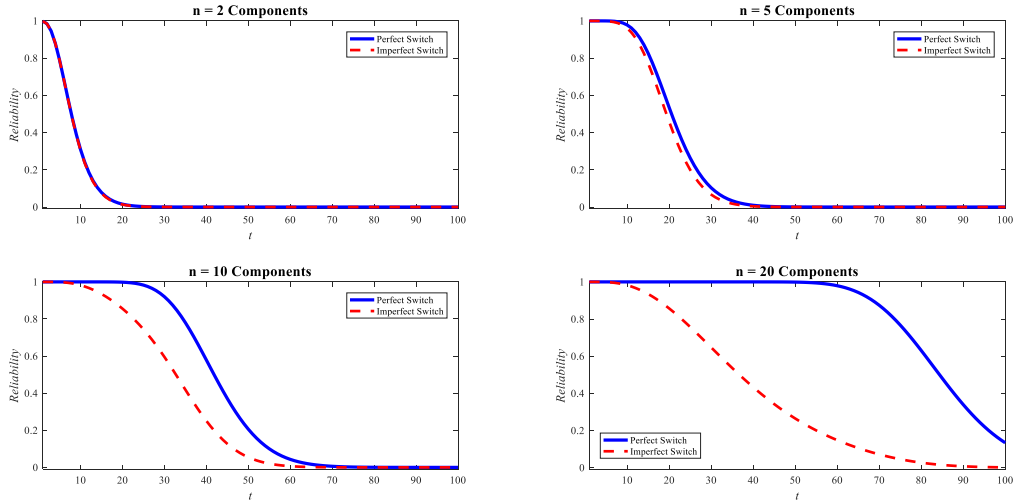

**Fig. 2.** Comparison of Reliability Function Graphs.

As observed, the system reliability with a perfect switch consistently outperforms that with an imperfect switch throughout the mission. Furthermore, as the number of system components increases, the reliability, particularly under a perfect switch, increases significantly.

For the sensitivity analysis, Eqs. (6) and (10) are examined. These equations contain six parameters: $\alpha, \beta, \theta, \gamma, \eta$ and $n$, where $n$ is the number of system components and is considered constant. Therefore, the influence of each of the remaining five parameters on the system reliability function and the *MTTF* is investigated. Fig. 3 illustrates the sensitivity analysis plot of the reliability function for each parameter.

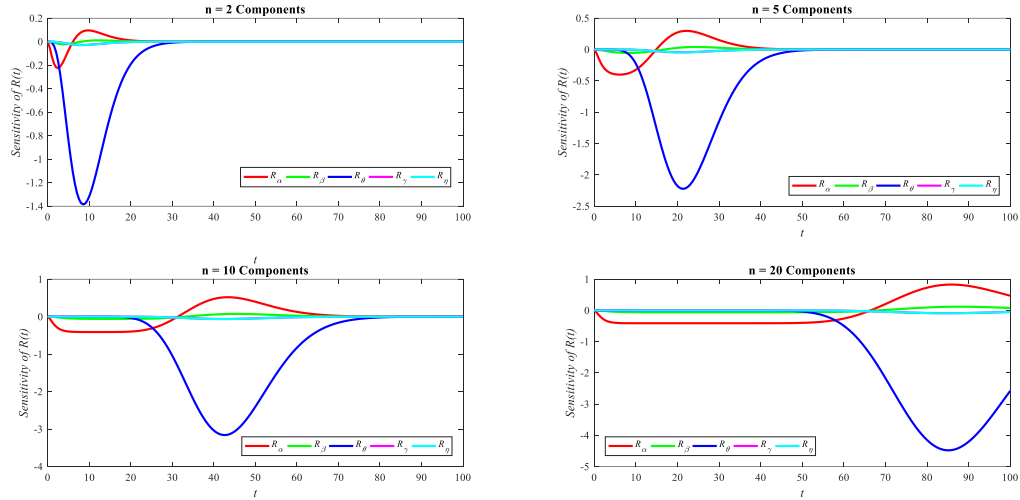

**Fig. 3.** Sensitivity Analysis of Reliability with Different Parameters.

According to Fig. 3, it is observed that the parameter with the greatest influence on the reliability function is the scale parameter of $\theta$, while the least influence is attributed to parameters $\eta$ and $\gamma$. The sensitivity analysis for the *MTTF* is provided in Tables 1 through 5.

Based on the mathematical analysis performed, parameters $\alpha$ and $\beta$ appear linearly in the *MTTF* formula. This characteristic causes the parameters themselves to be eliminated from the equation when calculating the partial derivatives with respect to them. In other words, the sensitivity of *MTTF* for $\alpha$ and $\beta$ is a function of the other system parameters. Consequently, the degree of influence of $\alpha$ and $\beta$ on *MTTF* is constant and predictable. Tables 1 and 2 demonstrate this finding.

**Table 1**

Sensitivity analysis of *MTTF* with $\alpha$.

| N | 2 | 5 | 10 | 20 |
|---|---|---|---|---|
| for each $\alpha$ | 3.4932 | 8.7329 | 17.4658 | 34.9317 |

**Table 2**

Sensitivity analysis of *MTTF* with $\beta$.

| N | 2 | 5 | 10 | 20 |
|---|---|---|---|---|
| for each $\beta$ | 0.5068 | 1.2671 | 2.5342 | 5.0683 |

In Tables 1 and 2, all values are observed to be positive, indicating a direct relationship with the MTTF. However, the values for parameter $\alpha$ are larger than their corresponding values for parameter $\beta$, implying that the effect of parameter $\alpha$ is stronger than that of parameter $\beta$.

Tables 3 through 5 also present the sensitivity analysis of the *MTTF* index with respect to parameters $\theta, \gamma$, and $\eta$.

**Table 3**

Sensitivity analysis of *MTTF* with $\theta$.

|  | n=2 | n =5 | n =10 | n =20 |
|---|---|---|---|---|
| $\theta$=0.3 | -43.3426 | -108.357 | -216.713 | -433.426 |
| $\theta$=0.4 | -23.9595 | -59.8988 | -119.798 | -239.595 |
| $\theta$=0.5 | -15.0662 | -37.6654 | -75.3309 | -150.662 |
| $\theta$=0.6 | -10.3115 | -25.7787 | -51.5573 | -103.115 |
| $\theta$=0.7 | -7.5094 | -18.7734 | -37.5468 | -75.0937 |

**Table 4**

Sensitivity analysis of *MTTF* with $\gamma$.

|  | n=2 | n =5 | n =10 | n =20 |
|---|---|---|---|---|
| $\gamma = 1.2$ | -0.4332 | -1.0829 | -2.1659 | -4.3318 |
| $\gamma = 1.5$ | -0.2951 | -0.7377 | -1.4754 | -2.9507 |
| $\gamma = 1.8$ | -0.2138 | -0.5346 | -1.0692 | -2.1385 |
| $\gamma = 2.2$ | -0.1489 | -0.3724 | -0.7447 | -1.4894 |
| $\gamma = 2.5$ | -0.1179 | -0.2947 | -0.5894 | -1.1787 |

**Table 5**

Sensitivity analysis of *MTTF* with $\eta$.

|  | n=2 | n=5 | n=10 | n=20 |
|---|---|---|---|---|
| $\eta = 1.5$ | -0.428 | -1.0699 | -2.1399 | -4.2798 |
| $\eta = 1.8$ | -0.3739 | -0.9348 | -1.8696 | -3.7393 |
| $\eta = 2.2$ | -0.3068 | -0.767 | -1.534 | -3.0679 |
| $\eta = 2.5$ | -0.2615 | -0.6537 | -1.3074 | -2.6149 |
| $\eta = 3.0$ | -0.1969 | -0.4922 | -0.9844 | -1.9687 |

As observed, all values in Tables 3 through 5 are negative, indicating an inverse relationship with the *MTTF*. In other words, as each of the three parameters increases, the sensitivity of the *MTTF* decreases. Parameter $\theta$ has the greatest effect on the system, while the least effect is attributed to parameters $\eta$ and $\gamma$.

## 6. Conclusion and Future work

This paper studied the reliability analysis of a 1-out-of-*n* cold standby redundant system where the time-to-failure of the system components and the switch follows the Generalized Lindley distribution. An explicit equation was derived for the case of a perfect switch. To provide this equation, the probability density function of the sum of *n* independent and identically distributed random variables from the Generalized Lindley distribution was required, which was obtained using the moment generating function. A lower bound was also calculated for the case of an imperfect switch. Furthermore, a sensitivity analysis was performed on the system reliability function. It was observed that the most critical parameter for the system was the scale parameter $\theta$.

One area for future work could involve redundancy allocation problems. Extensive research has been conducted in this area. The authors plan to model a system introduced in [26] using different methods and calculate its maximum reliability under constraints on the cost, weight and volume of the system components.

# References

[1] Xing, L., Tannous, O., & Dugan, J. B. (2011). Reliability analysis of nonrepairable cold-standby systems using sequential binary decision diagrams. *IEEE Transactions on Systems, Man, and Cybernetics-Part A: Systems and Humans*, *42*(3), 715-726.

[2] Wang, C., Xing, L., & Amari, S. V. (2012). A fast approximation method for reliability analysis of cold-standby systems. *Reliability Engineering & System Safety*, *106*, 119-126.

[3] Wang, C., Wang, X., Xing, L., Guan, Q., Yang, C., & Yu, M. (2021). A fast and accurate reliability approximation method for heterogeneous cold standby sparing systems. *Reliability Engineering & System Safety*, *212*, 107596.

[4] Coit, D. W. (2001). Cold-standby redundancy optimization for nonrepairable systems. *Iie Transactions*, *33*(6), 471-478.

[5] Billinton, R., & Allan, R. N. (1985). *Reliability evaluation of engineering systems* (Vol. 198, No. 3). New York: Pitman.

[6] Hoyland, A., & Rausand, M. (2009). *System reliability theory: models and statistical methods*. John Wiley & Sons.

[7] Dhillon, B. S. (2004). *Reliability, quality, and safety for engineers*. CRC press.

[8] Sadeghi, M., Roghanian, E., Shahriari, H., & Sadeghi, H. (2021). Reliability optimization for non-repairable series-parallel systems with a choice of redundancy strategies and heterogeneous components: Erlang time-to-failure distribution. *Proceedings of the institution of mechanical engineers, part o: journal of risk and reliability*, *235*(3), 509-528.

[9] Akhavan Niaki, S. T., & Yaghoubi, A. (2021). Exact equations for the reliability and mean time to failure of 1-out-of-n cold-standby system with imperfect switching. *Journal of Optimization in Industrial Engineering*, *14*(2), 197-203.

[10] Yaghoubi, A., & Moradi, N. (2023). An explicit formula for reliability of 1-out-of-n cold standby spare systems with Weibull distribution. *International Journal of Reliability, Quality and Safety Engineering*, *30*(01), 2250029.

[11] Lindley, D. V. (1958). Fiducial distributions and Bayes' theorem. *Journal of the Royal Statistical Society. Series B (Methodological)*, 102-107.

[12] Ghitany, M. E., Atieh, B., & Nadarajah, S. (2008). Lindley distribution and its application. *Mathematics and computers in simulation*, *78*(4), 493-506.

[13] Shanker, R., & Mishra, A. (2013). A quasi Lindley distribution. *African Journal of Mathematics and Computer Science Research*, *6*(4), 64-71.

[14] Zakerzadeh, H., & Dolati, A. (2009). Generalized lindley distribution.

[15] Nadarajah, S., Bakouch, H. S., & Tahmasbi, R. (2011). A generalized Lindley distribution. *Sankhya B*, *73*(2), 331-359.

[16] Abouammoh, A. M., Alshangiti, A. M., & Ragab, I. E. (2015). A new generalized Lindley distribution. *Journal of Statistical computation and simulation*, *85*(18), 3662-3678.

[17] Kantar, Y. M., Usta, I., Arik, I., & Yenilmez, I. (2018). Wind speed analysis using the extended generalized Lindley distribution. *Renewable Energy*, *118*, 1024-1030.

[18] Ahsan-ul-Haq, M., Choudhary, S. M., AL-Marshadi, A. H., & Aslam, M. (2022). A new generalization of Lindley distribution for modeling of wind speed data. Energy Reports, 8, 1-11.

[19] Barlow, R. B. (1975). Statistical theory of reliability and life testing. Holt.

[20] Marshall, A. W., Olkin, I., & Arnold, B. C. (1979). Inequalities: theory of majorization and its applications.

[21] Yen, T. C., Chen, W. L., & Chen, J. Y. (2016). Reliability and sensitivity analysis of the controllable repair system with warm standbys and working breakdown. *Computers & Industrial Engineering*, *97*, 84-92.

[22] Manglik, M., & Ram, M. (2013). Reliability analysis of a two unit cold standby system using Markov process. *Journal of Reliability and Statistical Studies*, 65-80.

[23] Naaz, S., Ram, M., & Kumar, A. (2025). Reliability and Sensitivity Appraisal of Information Processing Server System: Structure-Function Approach. *ASCE-ASME J Risk and Uncert in Engrg Sys Part B Mech Engrg*, 1-23.

[24] Birolini, A. (2010). Reliability engineering: theory and practice. Berlin, Heidelberg: Springer Berlin Heidelberg.

[25] Koren, I., & Krishna, C. M. (2007). *Fault-Tolerant Systems*. Morgan Kaufmann Publishers.

[26] Fyffe, D.E., Hines, W.W. and Lee, N.K. (1968) System reliability allocation and a computational algorithm. *IEEE Transactions 011 Reliability,* R-I7, 64-69.